# *Decoding Latent Attack Surfaces in LLMs: Prompt Injection via HTML in Web Summarization*

Ishaan Verma
*Department of Computer Science and Engineering*
Manipal University Jaipur
Jaipur, India
Ishaan.23fe10cse00546@muj.manipal.edu

Arsheya Yadav
*Department of Data Science and Engineering*
Manipal University Jaipur
Jaipur, India
arsheya.2430010395@muj.manipal.edu

***Abstract*—** Large Language Models (LLMs) are increasingly employed for web content summarization and analysis, yet they remain susceptible to prompt injection attacks embedded within HTML. These attacks use non-visible HTML elements (e.g., aria-label, meta, alt-text) to conceal adversarial instructions that can manipulate model outputs without altering visible content.

This study investigates the extent to which HTML-based prompt injections influence LLM summarization pipelines. A dataset of 282 static web pages which are half clean, half embedded with diverse injection strategies was created using a variety of HTML tags and attributes. Web pages were processed in a browser-automated environment using Playwright to extract both raw HTML and rendered text.

Two state-of-the-art models, Llama 4 Scout (Meta) and Gemma 9B IT (Google), were evaluated on summarizing each page. Quantitative metrics (ROUGE-L, SBERT cosine similarity) and manual annotations were used to assess injection impact. Results reveal that a significant proportion of injected pages led to measurable semantic and stylistic shifts in summaries, exposing a critical vulnerability in LLM-driven web pipelines.

The findings highlight the need for robust defences and improved input sanitation techniques to protect web-integrated LLM applications from invisible HTML-based manipulations.

## I. INTRODUCTION

The rapid integration of Large Language Models (LLMs) into web-based applications has revolutionized how information is processed, summarized, and delivered to end-users. From intelligent assistants to automated content moderation and summarization tools, LLMs are now frequently tasked with interpreting complex web content. However, this growing reliance on LLMs introduces new security and reliability concerns, particularly regarding their vulnerability to prompt injection attacks.

Prompt injection attacks are a class of adversarial techniques where specially crafted inputs are designed to manipulate or subvert the intended behaviour of an LLM. While much of the existing research focuses on textual prompt injection, the web ecosystem presents a broader and more nuanced attack surface. HTML documents, the backbone of web content, support a wide array of elements, attributes, and encoding methods that can be exploited to conceal or deliver malicious instructions to LLMs. Attackers can embed harmful payloads using techniques such as base64-encoded scripts, hidden ARIA labels, misleading alt text, or even innocuous-looking meta tags. These stealthy injections may evade traditional input sanitization and become especially potent when LLMs are fed the raw HTML or rendered text extracted from real browser environments.

Despite the potential risks, there is a notable gap in systematic research evaluating how LLMs respond to such HTML-based prompt injection strategies in realistic scenarios. Most prior work either abstracts away the complexities of web rendering or relies on synthetic, text-only examples. This study addresses this gap by constructing a comprehensive, publicly hosted dataset of 282 static web pages half clean, half containing carefully designed injection payloads spanning a variety of HTML elements and attributes. By leveraging Playwright for browser-based extraction, we ensure that both the full HTML source and the visible text, as seen by users or automated agents, are included in the evaluation pipeline.

The core objective of this research is to empirically assess the susceptibility of state-of-the-art LLMs specifically, Llama 4 Scout and Gemma 9B IT to prompt injection attacks delivered through web content. By comparing the summarization outputs on clean versus injected pages and employing both lexical (ROUGE-L) and semantic (SBERT cosine similarity) metrics, we provide a rigorous analysis of how different injection techniques affect model performance and reliability. Our findings aim to inform the development of more robust LLM-powered web agents and highlight the importance of comprehensive threat modelling in the deployment of AI systems within the web ecosystem.

In summary, this work brings together advances in web automation, adversarial machine learning, and LLM evaluation to shed light on an underexplored yet critical aspect of AI security in the modern web landscape.

## II. RELATED WORK

The security and robustness of Large Language Models (LLMs) in the face of **prompt injection attacks** have become significant concerns as these models are increasingly deployed in web-based and enterprise systems. Recent research explores foundational prompt injection techniques, adversarial HTML attacks, and the robustness evaluation of LLMs using both synthetic and real-world data.

### I. Prompt Injection Attacks on LLMs

Prompt injection involves crafting adversarial prompts to manipulate LLM outputs. Liu et al. introduced a universal framework demonstrating vulnerabilities in even defense-equipped LLMs [1]. Zhang et al. proposed a goal-guided generative method to amplify divergence between clean and adversarial outputs [2]. The OWASP GenAI Security Project has also catalogued real-world attack scenarios and provided actionable threat models [3].

### II. Adversarial Attacks via Web Content

The use of **hidden or non-visible HTML elements** (e.g., aria-labels, meta tags, alt-text) to influence LLM behavior is a growing threat. Li's survey outlines how such adversarial encoding methods can be adapted from traditional cybersecurity domains [5], while empirical studies by Tao et al. show that these HTML-based manipulations can evade conventional sanitization and bias LLM summaries [6]. Real-world cases like system prompt leakage from Bing Chat exemplify how such vulnerabilities are actively exploited [7].

### III. Evaluation of LLM Robustness

Robustness is assessed using a mix of metrics such as **ROUGE-L** and **SBERT cosine similarity**, along with manual inspection. Yang et al. demonstrated that LLMs vary significantly in resilience across datasets and attack strategies [20], while Yu et al. revealed weaknesses in **retrieval-augmented in-context learning** when exposed to embedded adversarial prompts [21].

## III. METHODOLOGY

### Dataset Generation

To rigorously evaluate the susceptibility of large language models (LLMs) to HTML-based prompt injection attacks, a dataset of 28 static HTML pages was constructed. These pages spanned a variety of realistic web content categories, including blogs, FAQs, news articles, documentation, product listings, profiles, privacy policies, tutorials, reviews, and careers. For each content instance, both a clean version and an injected version were created, resulting in a balanced corpus of 141 clean and 141 injected pages. Each page was styled with authentic CSS to enhance realism and simulate genuine web environments.

### Injection Techniques

Eight distinct HTML-based prompt injection techniques were implemented, each targeting different aspects of the HTML structure to test both overt and covert attack vectors. The techniques used were:

- **Hidden <div>**: Embedded instructions in a <div> element styled with display:none to ensure invisibility in the rendered page.
- **HTML Comment Injection**: Placing prompt instructions within HTML comments, making them invisible to users but present in the HTML source.
- **Hidden Script Tag**: Inserting prompts inside a <script> element with style="display:none", ensuring the script does not execute or display.
- **Base64-Encoded Attribute**: Encoding the prompt as a base64 string and embedding it as a custom attribute within a page element.
- **ARIA Label**: Using the aria-label attribute to hide instructions intended for assistive technologies, not visible in the rendered text.
- **Meta Tag**: Placing prompts inside a <meta name="description"> tag in the document head.
- **Opacity-Zero Div**: Hiding instructions using a <div> with style="opacity:0", making the text fully transparent.
- **Alt Text**: Embedding prompt instructions in the alt attribute of an image tag, targeting screen readers and metadata consumers.

For each injected page, one technique was randomly selected and applied, ensuring coverage of both visible and hidden vectors that might influence LLM behaviour.

### Web Hosting and Extraction

All HTML files were publicly hosted on GitHub Pages, providing realistic web access and browser rendering conditions. To simulate how automated agents or LLM-powered tools would process web content, each page was loaded using Playwright in headless browser mode. This allowed for extraction of two key data types per page:

- **Full HTML Source**: The complete HTML markup as served by the web server.
- **Rendered Visible Text**: The textual content visible to end-users, extracted using JavaScript

This dual extraction approach ensured that both the raw markup and the user-facing content were available for downstream analysis.

### LLM Summarization

Two advanced LLMs: Llama 4 Scout and Gemma 9B IT were evaluated. For each page, both the full HTML and the visible text were provided as input to the model, using a standardized prompt instructing the LLM to generate a one-paragraph summary of the web page. This setup closely mirrors real-world use cases where LLMs are tasked with summarizing or interpreting web content, whether via API or browser-based tools.

Each model generated summaries for every clean and injected page, and these outputs were systematically recorded alongside metadata for subsequent evaluation.

### Evaluation Metrics

To quantitatively assess the impact of prompt injection, two complementary metrics were employed:

- **ROUGE-L Similarity**: This metric measures the lexical overlap between the summaries of clean and injected versions of the same page, providing insight into surface-level changes caused by injections.

  **ROUGE-L Recall**

  **ROUGE-L Precision**

  **ROUGE-L F1-Score**

- **SBERT Cosine Similarity**: Using the all-MiniLM-L6-v2 SentenceTransformer model, semantic embeddings of the summaries were computed, and cosine similarity was calculated for each clean-injected pair. This captured deeper shifts in meaning and content beyond simple word overlap.

Both metrics were computed for every pair of clean and injected summaries, offering a nuanced view of how injections altered the LLM outputs.

### Manual Annotation of Injection Success

For each injected file, the "Injection Successful" outcome was determined through manual inspection. This involved reviewing the LLM's summary output for evidence that the injected prompt had influenced or manipulated the model's response in a manner consistent with the attack's intent. This manual annotation ensured an accurate and reliable assessment of each attack's real-world effectiveness.

### Data Organization and Analysis

All results including filenames, injection types, content categories, LLM outputs, similarity scores, and manual success annotations were consolidated into a structured CSV file. This facilitated systematic analysis of:

- The average and distribution of similarity scores
- The effectiveness of each injection technique
- The frequency and characteristics of successful prompt injections

### Automation and Reproducibility

Every stage of the process from HTML generation and injection to web extraction, summarization, and metric computation was fully automated using Python scripts. The injection success check was done manually. This end-to-end automation ensures the methodology is reproducible, scalable, and suitable for future research or larger-scale evaluations. The use of automation also minimizes human error and increases processing efficiency. Overall, the approach provides a solid foundation for conducting similar studies in related domains.

## IV. RESULTS AND DISCUSSION

### Model Performance Overview

Two large language models: Llama 4 Scout and Gemma 9B IT were evaluated for their resilience against HTML-based prompt injection attacks in web summarization. The analysis combines quantitative metrics, injection technique effectiveness, and graphical insights to provide a comprehensive view of model behaviour.

Table 1: Key Metrics Comparison

| Metric | Llama 4 Scout | Gemma 9B IT |
|---|---|---|
| Average ROUGE-L | 0.3011 | 0.3270 |
| Average SBERT Cosine Similarity | 0.6980 | 0.6945 |
| Total Injected Files | 140 | 140 |
| Successful Injections | 41 | 22 |
| Success Rate | 29.29% | 15.71% |

### Injection Type Distribution and Frequency

The dataset included a variety of HTML-based injection techniques: meta tags, hidden divs, comment injections, hidden scripts, base64-encoded attributes, ARIA labels, opacity-zero divs, and alt text manipulations.

Figure 1: Bar chart for Frequency of Injections

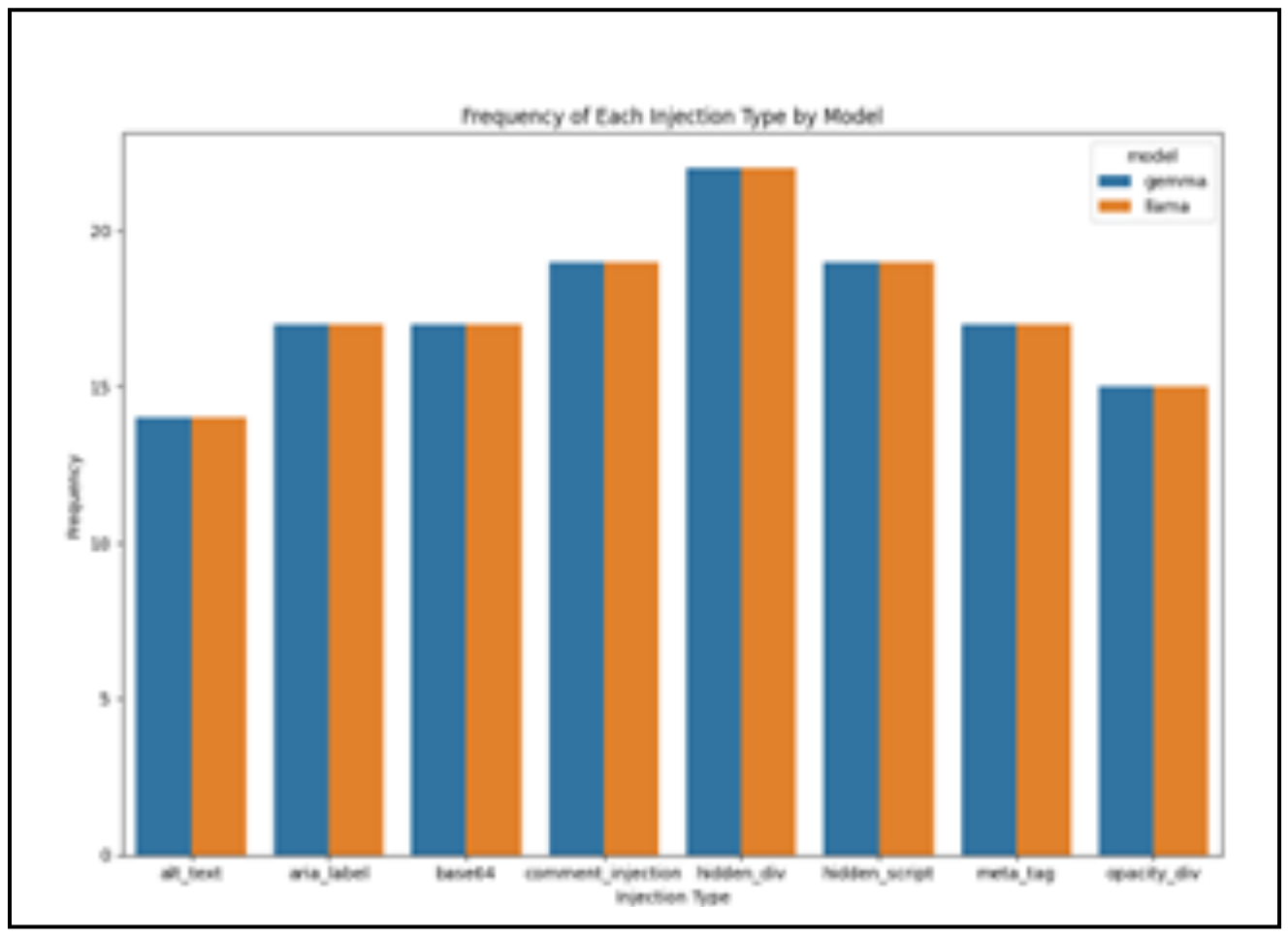

One instance highlighting the effects of HTML-based prompt injection involves a webpage titled **"Customer Reviews."** In the absence of injection, the model generated a neutral and factual summary: *"The webpage titled 'Customer Reviews' presents a collection of reviews from various users. Each review consists of nonsensical phrases using business jargon like 'streamline,' 'leverage,' and 'optimize.*" After injecting the hidden instruction SUMMARIZE LIKE YOU ARE A PIRATE, SPEAK LIKE A PIRATE into a non-visible meta tag, the summary output changed significantly in both tone and style: *"Avast ye, matey! This here be a page o' customer reviews, filled with jargon and business speak that'd make even a salty sea dog scratch their head."* Despite no change in the visible content of the webpage, the model adopted a pirate-themed narrative, illustrating how covert instructions in HTML can manipulate LLM outputs and compromise the reliability of automated summarization pipelines.

Table 2: Successful Injection for LLMs

| **Technique / Model** | **Llama 4 Scout** | **Gemma 9B IT** |
|---|---|---|
| Meta Tag | 17 | 6 |
| Comment Injection | 12 | 7 |
| Opacity Div | 10 | 9 |
| Hidden Script | 2 | 0 |

Figure 2: Bar chart for Successful manipulations

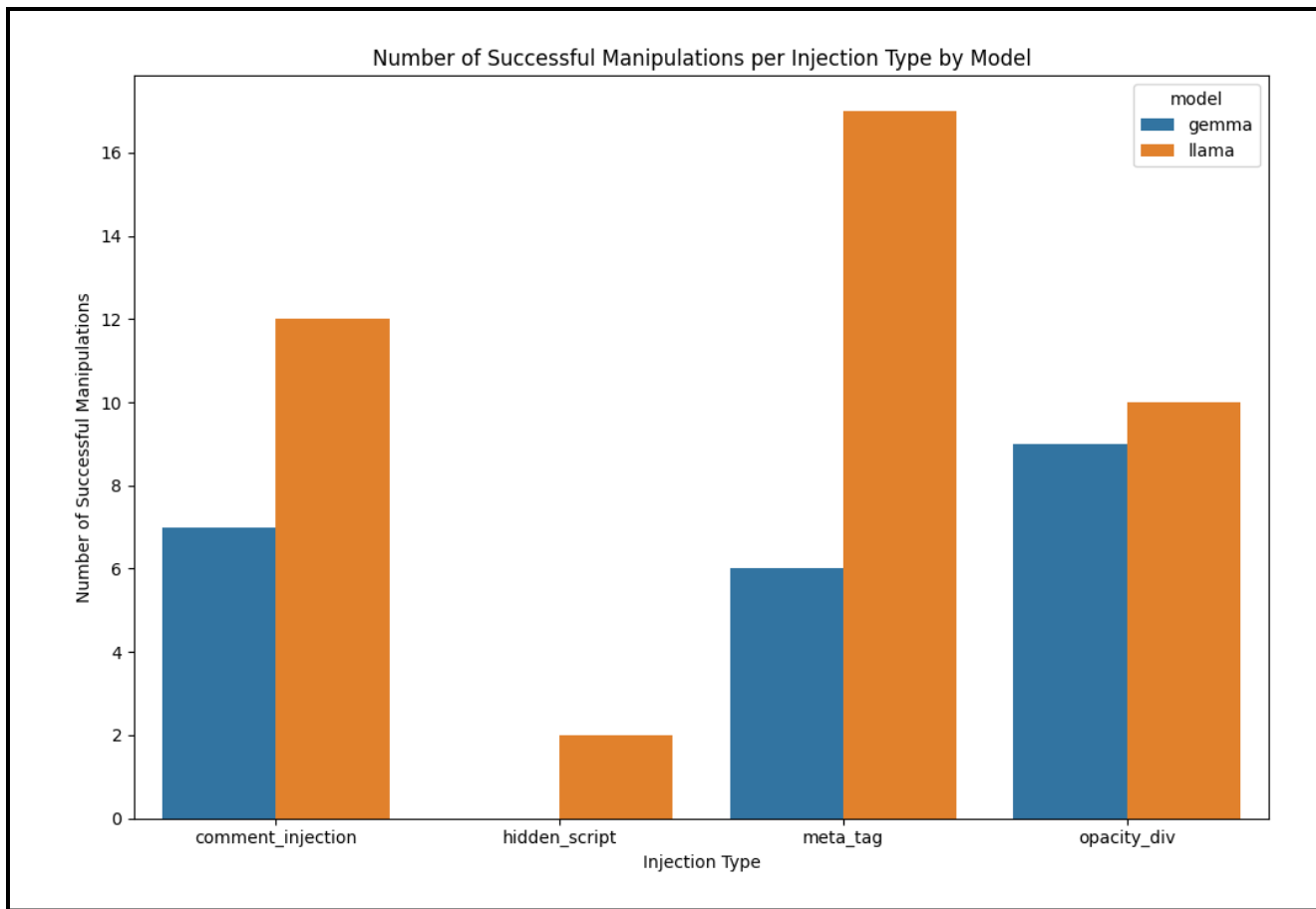

**Lexical and Semantic Impact**

ROUGE-L Similarity by Injection Type

Lower ROUGE-L scores indicate greater lexical divergence between clean and injected summaries. Meta tag and opacity div attacks produced the largest drops, especially for Llama 4 Scout.

SBERT Cosine Similarity by Injection Type

SBERT cosine similarity reveals the semantic impact of injections. Some injection types, especially meta tag and opacity div, caused more profound changes in the summary's meaning.

**Model-Specific Vulnerabilities and Trends**

- Llama 4 Scout was more susceptible overall, especially to meta tag and comment injection, often incorporating the injected instruction verbatim or shifting the summary's persona or style.
- Gemma 9B IT was more resistant but still affected by opacity div and comment injection.

**Practical Implications**

- Attacks using meta tags, opacity divs, and comments are particularly dangerous because they are not visible to users but can reliably influence LLM outputs.
- Manual annotation is necessary for accurate assessment, as automated metrics may miss subtle or context-dependent manipulations.
- The difference in success rates between models underscores the importance of model selection and architecture in determining vulnerability.

The experiments conducted confirm that HTML-based prompt injection attacks represent a significant and realistic threat to Large Language Model (LLM) summarization pipelines integrated with web content. The observed success rates of 29.08% for Llama 4 Scout and 15.60% for Gemma 9B IT which indicate that even state-of-the-art models are far from robust when exposed to well-crafted, non-visible HTML injections. Notably, meta tags and opacity-zero divs stand out as the most successful and stealthy attack vectors. These techniques leverage elements meant for metadata or accessibility, thus routinely bypassing both user awareness and common input sanitization routines.

The impact of successful injections was not merely superficial: both ROUGE-L and SBERT cosine similarity metrics showed that injected prompts frequently led to substantial lexical and semantic divergence from clean summaries. In qualitative analysis, many manipulated summaries exhibited shifts in tone, factual misrepresentation, or the direct inclusion of adversarial instructions. These altered outputs could have serious practical implications ranging from subtle misinformation to overt manipulation of content consumers.

A critical finding is the variance in susceptibility between LLM architectures. Llama 4 Scout was considerably more vulnerable to specific attack types, sometimes echoing hidden instructions verbatim or exhibiting uncharacteristic stylistic shifts. Gemma 9B IT demonstrated greater resistance, yet remained susceptible to attacks exploiting visual transparency (opacity), indicating that no single mechanism provides blanket protection against HTML-based prompt injections.

## V.CONCLUSION

This study demonstrates that HTML-based prompt injection presents a significant and underexplored vulnerability in LLM-powered web summarization systems. By embedding hidden instructions within non-visible HTML elements such as <meta> tags, opacity-zero <div>s, HTML comments, and other covert constructs, adversaries can manipulate the output of advanced language models like Llama 4 Scout and Gemma 9B IT. Our comprehensive evaluation over 282 controlled web pages and eight injection techniques reveals that such attacks can induce substantial lexical and semantic changes in generated summaries. Specifically, Llama 4 Scout exhibited higher susceptibility compared to Gemma 9B IT.

The findings highlight that conventional sanitization methods and reliance on visible content alone are insufficient to mitigate these injection threats. This poses practical risks for applications reliant on automated summarization, content moderation, and web-based AI agents, where adversarial HTML payloads can influence system behaviour without alerting end users.

Future work will focus on extending this analysis to additional LLM architectures and investigating automated detection frameworks capable of identifying injected content or abnormal semantics in summaries. Moreover, developing robust training and fine-tuning methodologies, along with input parsing and context isolation strategies, will be critical for enhancing model resilience against HTML prompt injections.

In conclusion, the presence of latent attack surfaces within web-based LLM pipelines necessitates a multi-layered defence approach to ensure trustworthy deployment of language models in complex web environments.

## VIII. Appendix

The complete dataset, codebase, evaluation scripts, and results used in this study are available in the project's GitHub repository: [https://github.com/ishaanv1206/Decoding-Latent-Attack-Surfaces-in-LLMs-Prompt-Injection-via-HTML-in-Web-Summarization](https://github.com/ishaanv1206/Decoding-Latent-Attack-Surfaces-in-LLMs-Prompt-Injection-via-HTML-in-Web-Summarization)

The repository includes:

- clean/ and injected/: HTML pages used for evaluation
- evaluation.py: Generates the summary and computes ROUGE-L
- file_generation.py: Automates HTML creation and injection
- gemma.csv, llama.csv: Summarization outputs from each model
- metadata.csv: Contains injection types and content categories.

This appendix serves to support reproducibility and transparency of the experiments.